\documentclass[aps,prb,twocolumn,superscriptaddress,showpacs]{revtex4-1}%,preprint,groupedaddress
\usepackage{amsmath}
\usepackage{graphicx}

\begin{document}

\title{Vectorial spin-polarization detection in multichannel spin-resolved photoemission spectroscopy using an Ir(001) imaging spin filter}

\author{Erik D. Schaefer}
\affiliation{Institut f{\"u}r Physik, Johannes Gutenberg-Universit{\"a}t Mainz, Staudingerweg 7, 55128 Mainz, Germany}
\author{Stephan Borek}
\author{J{\"u}rgen Braun}
\affiliation{Department Chemie, Ludwig-Maximilians-Universit{\"a}t M{\"u}nchen, Butenandtstraße 5-13, 81377 M{\"u}nchen, Germany}
\author{J{\'a}n Min{\'a}r}
\affiliation{Department Chemie, Ludwig-Maximilians-Universit{\"a}t M{\"u}nchen, Butenandtstraße 5-13, 81377 M{\"u}nchen, Germany}
\affiliation{New Technologies-Research Centre, University of West Bohemia, Univerzitni 8, 306 14 Pilsen, Czech Republic}
\author{Hubert~Ebert}
\affiliation{Department Chemie, Ludwig-Maximilians-Universit{\"a}t M{\"u}nchen, Butenandtstraße 5-13, 81377 M{\"u}nchen, Germany}
\author{Katerina Medjanik}
\author{Gerd Sch{\"o}nhense}
\author{Hans-Joachim Elmers}
\affiliation{Institut f{\"u}r Physik, Johannes Gutenberg-Universit{\"a}t Mainz, Staudingerweg 7, 55128 Mainz, Germany}

\date{\today}

\begin{abstract}
We report on spin- and angular resolved photoemission spectroscopy using a high-resolution imaging spin filter based on a large Ir(001) crystal enhancing the effective figure of merit for spin detection by a factor of over $10^3$ compared to standard single channel detectors. Furthermore, we review the spin filter preparation, and its lifetime. The spin filter efficiency is mapped on a broad range of scattering energies and azimuthal angles. Large spin filter efficiencies are observed for the spin component perpendicular as well as parallel to the scattering plane depending on the azimuthal orientation of the spin filter crystal. A spin rotator capable of manipulating the spin direction prior to detection complements the measurement of three observables, thus allowing for a derivation of all three components of the spin polarization vector in multichannel spin polarimetry. The experimental results nicely agree with spin-polarized low energy electron diffraction calculations based on a fully relativistic multiple scattering method in the framework of spin-polarized density functional theory.
\end{abstract}


\maketitle
\section{Introduction}%\section{\label{}}
Over the past decades, photoemission spectroscopy (PES) has been a key technique for the investigation of electronic properties in solid state materials \cite{SUG15}. In recent years, many promising materials such as topological insulators \cite{HAS10,KOE07}, metal-organic interfaces \cite{KAH03} or Heusler compounds \cite{FEL15} have been studied by photoemission. For these materials, it is important to analyze spin-resolved spectra in order to understand their unique properties \cite{NIS10,DJE16,JOU14}. So far, the lack of measurement efficiency in spin-resolved experiments combined with a short lifetime of samples due to radiation damages or surface contamination often prohibited the analysis of such sensitive materials.\\
PES enjoyed a considerable increase in performance due to parallelization concepts as used in energy- and angle-resolved photoemission spectroscopy (ARPES). However, the spin-polarization analysis remained time consuming. The most widely used spin filtering techniques are based on Mott scattering\cite{KLI66}, spin-polarized low-energy electron diffraction (SPLEED)\cite{KIR81}, or very-low-energy electron diffraction (VLEED)\cite{TIL89,OKU09}. Combined with PES these techniques comprise single channel detectors and therefore suffer from low efficiency.\\
Recently, a spin-resolved multichannel technique has been introduced \cite{KOL11,TUS11}. The effective figure of merit value (FoM$_{2D}$) which reflects the overall filtering performance has been enhanced by up to 3 orders of magnitude \cite{KOL11}. Multichannel spin detection thus opens a pathway to study sensitive materials requiring fast measurements with high measurement efficiency. In the following years, the multichannel concept has been further developed \cite{TUS11,STR15,JI16,KUT16}.\\
A present drawback of parallel (imaging) spin detectors is the fact that one detects only a single spin component without changing the experimental geometry. However, a measurement of more than one vector component of the spin polarization is of high interest in particular for topological surface states exhibiting a complex spin structure. Striving for improving this situation, we report on an investigation of the azimuthal rotation of the Ir(001) spin filter crystal with respect to the fixed scattering plane. In case of a coincidence of scattering plane and mirror plane the spin filter crystal exclusively detects the spin component perpendicular to the scattering plane. If the scattering plane deviates from the mirror plane the spin filter will become sensitive for the spin component parallel to the scattering plane. This approach thus allows for the sequential measurement of two independent components of the spin polarization vector. Applying a longitudinal magnetic field in order to rotate the transversal spin component prior to scattering further allows distinguishing longitudinal from transversal spin components. Thus, all three components of the spin polarization vector can be measured with the same setup.\\
The spin detection efficiency has been characterized for a wide range of scattering energies and azimuthal angles. The expected symmetry relations are confirmed.  We have identified parameters for maximum spin filter efficiency. Experimental results are interpreted by calculations that have been performed using fully relativistic multiple scattering techniques in the framework of spin-polarized density functional theory \cite{EBE11,EBE12}.
\section{Experiment}
\subsection{Experimental Setup}
\begin{figure}
	\centering
  \includegraphics[width=0.39\textwidth]{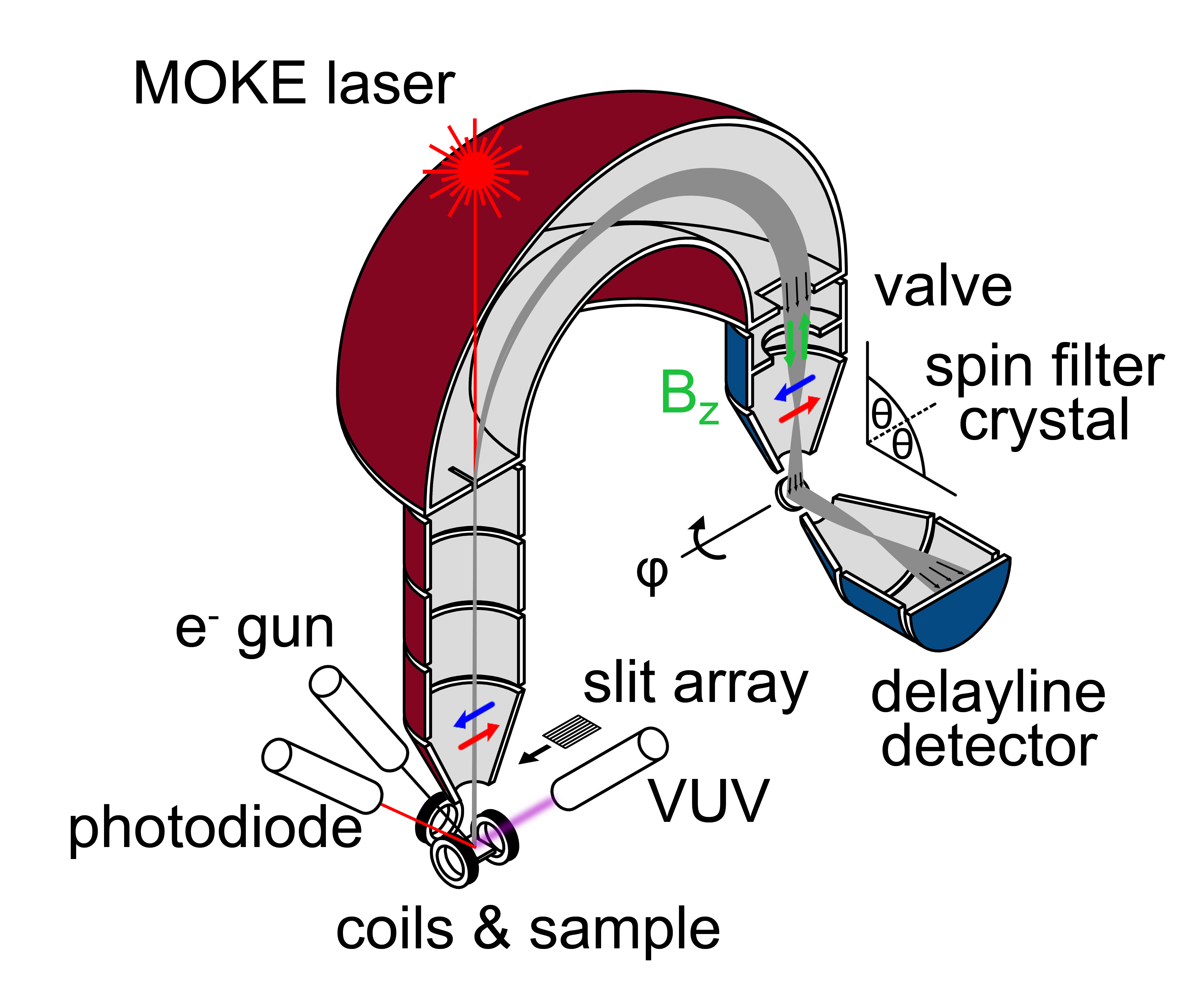}
	\caption{Experimental setup showing a cross section of the transfer lens and hemispherical analyzer (red) and spin filter part (blue). The out-of-plane spin component orientation is marked by red and blue arrows. The longitudinal magnetic field for the spin-polarization manipulation is marked by green arrows. Electron lenses schematic}
	\label{fig:analyzer}
\end{figure}

The experimental setup is shown in figure \ref{fig:analyzer}. The multichannel spin-, energy- and angle-resolving photoemission spectrometer is based on a commercial SPECS PHOIBOS 150 hemispherical analyzer equipped with the multi-element, two-stage transfer lens \cite{SPE12}. It is embedded in a $\mu$-metal main chamber where the W(110) substrate is mounted on a manipulator stage (base pressure $3\cdot10^{-10}$~mbar). Either the ($\overline{\Gamma \mathrm{N}}$)/$[1\overline{1}0]$ or the ($\overline{\Gamma \mathrm{H}}$)/$[001]$ axis is aligned along the angular dispersive direction of the hemispherical analyzer.\\
A longitudinal external magnetic field in front of the spin detector has been used for rotation of the spin polarization vector within the plane perpendicular to the beam axis.\\
The $\mu$-metal L-shaped spin filter chamber (base pressure $1\cdot10^{-10}$~mbar) with carbon coated electron optics is attached to the exit flange of the hemispherical analyzer. Moreover, a 5$^{\circ}$ inclination of the spin filter system is integrated to better match the simulated escape angles at the hemispherical analyzer exit plane. A metal plate valve is installed in the lens system between the exit plane and the scattering crystal to separate both chambers during crystal preparations. The installed internal $\mu$-metal shield near this valve is slightly permeable to inject the longitudinal magnetic field. Inside the spin filter chamber a circular Ir(001) crystal (\O = 15~mm) is destined for spin-dependent electron scattering based on SPLEED. The scattering angle $\theta$ is fixed to 45$^{\circ}$ and the spin filter crystal is mounted on a rotation feedthrough to vary the azimuthal angle $\varphi$. A bias voltage serves to vary the scattering energy. After reflection, the electron beam is imaged on a delayline detector offering spatial resolution and a high signal to background ratio.\\
For the angular calibration and resolution determination, a slit array ($d$ = 0.2~mm with 1~mm interval) has been moved in front of the entrance lens. The hemispherical analyzer has an additional built-in slit array at the exit plane generating equidistant lines in the energy dispersive direction ($d$ = 0.3~mm with 8~mm interval).\\
\subsection{Source of spin-polarized electrons}
For the characterization of the spin filter we used secondary electrons excited by an incident high energy (1.5~keV) electron beam from remanently magnetized epitaxial Fe/W(110) films with a defined direction of the spin polarization vector given by the magnetization direction of the sample. The easy magnetization axis depends on the film thickness. The magnetization direction points along $[1\overline{1}0]$ for thin films and changes to $[001]$ for film thicknesses larger than 6~nm. Photoemitted electrons from these films are spin-polarized along a direction parallel to the magnetization axis. The magnetic state of the Fe/W(110) sample has been analyzed by the longitudinal and transversal magneto-optical Kerr effect. A magnetization reversal of the Fe/W(110) sample induced by a magnetic field pulse leads to a beam polarization inversion of the secondary electrons. The polarization of secondary electrons from the Fe/W(110) shows maximum values in the kinetic energy range of 0-4~eV of approximately 45\% at primary beam energies of 0.8-2.0~kV (see Refs. \onlinecite{MIE11,KIR92,ALL86}).
\subsection{Substrate and spin filter preparation}
The W(110) substrate has been cleaned by 5-10 low-power flashes to 1200~K (corresponding to 75~W heating power) for 10~s (60~s idle interval) in an oxygen atmosphere of 5-8$\cdot10^{-8}$~mbar. After closing the oxygen supply and returning to base pressure, a subsequent high-power flash to 2200~K (150-180~W heating power) for 8-10~s results in a clean surface. The cleanness of W(110) has been confirmed using low energy electron diffraction (LEED).\\
Fe (thickness 3~nm) has been deposited at room temperature onto the tungsten crystal using molecular beam epitaxy. During the evaporation at rates of ca. 5~nm per hour the pressure rises to 5-10$\cdot10^{-10}$~mbar. A subsequent continuous annealing of the Fe/W(110) sample to 550-600~K results in a smooth surface as confirmed by LEED.\\
The Ir(001) spin filter crystal has been cleaned by 5-10 low-power flashes at 1200~K (135~W) for 10~s ($\ge$~5~min idle interval) in an oxygen atmosphere of $8\cdot10^{-8}$~mbar. After reaching the base pressure and shortly before starting a measurement, a high-power flash (1500-1600~K, 235~W) for 10~s desorbs the remaining oxygen (see also Refs.~\onlinecite{KIR13,TUS15,VAS15}).
\subsection{Electron optics}
The electron optical setup has been adjusted with the help of trajectory calculations using SIMION 8.1. The lens system images the exit plane of the hemispherical analyzer on the detector with an intermediate real image on the spin filter crystal. The scattering energy $E_{scatt} = E_{kin} + eV_{bias}$ has been varied by applying a bias voltage $V_{bias}$ to the spin filter crystal.
\section{Theory}
The calculations have been performed using the SPR-KKR program package \cite{EBE12}. It includes a spin-polarized relativistic version of the Korringa-Kohn-Rostoker (KKR) multiple scattering formalism. The self-consistent calculation of the atomic potentials necessary for the computation of the SPLEED spectra have been done using the tight-binding version of the KKR. This method is known for an effective description of surface systems. With the potentials the single-site scattering matrices are determined which are included later on in the determination of the Kambe X-matrix \cite{kambe}. The SPLEED spectra are simulated via the so-called layered-KKR method \cite{feder2,Pendry_SPLEED}. In this method the multiple scattering within the specific layers is considered via the above mentioned Kambe X-matrix. Additionally the scattering in between the atomic layers has to be considered. With both it is possible to determine the bulk reflection matrix which includes all information necessary for the simulation of the diffracted SPLEED intensities \cite{Pendry_SPLEED}. For the transition from the surface to vacuum one has to consider the image potential including a specific curve shape. We applied the Rundgren-Malmstr\"om barrier which is known for a successful description of the surface-barrier transition \cite{rundgren1,PhysRevB.87.195411}.
\section{Experimental Results}
\subsection{Resolution}
\begin{figure}
	\centering
  \includegraphics[width=0.45\textwidth]{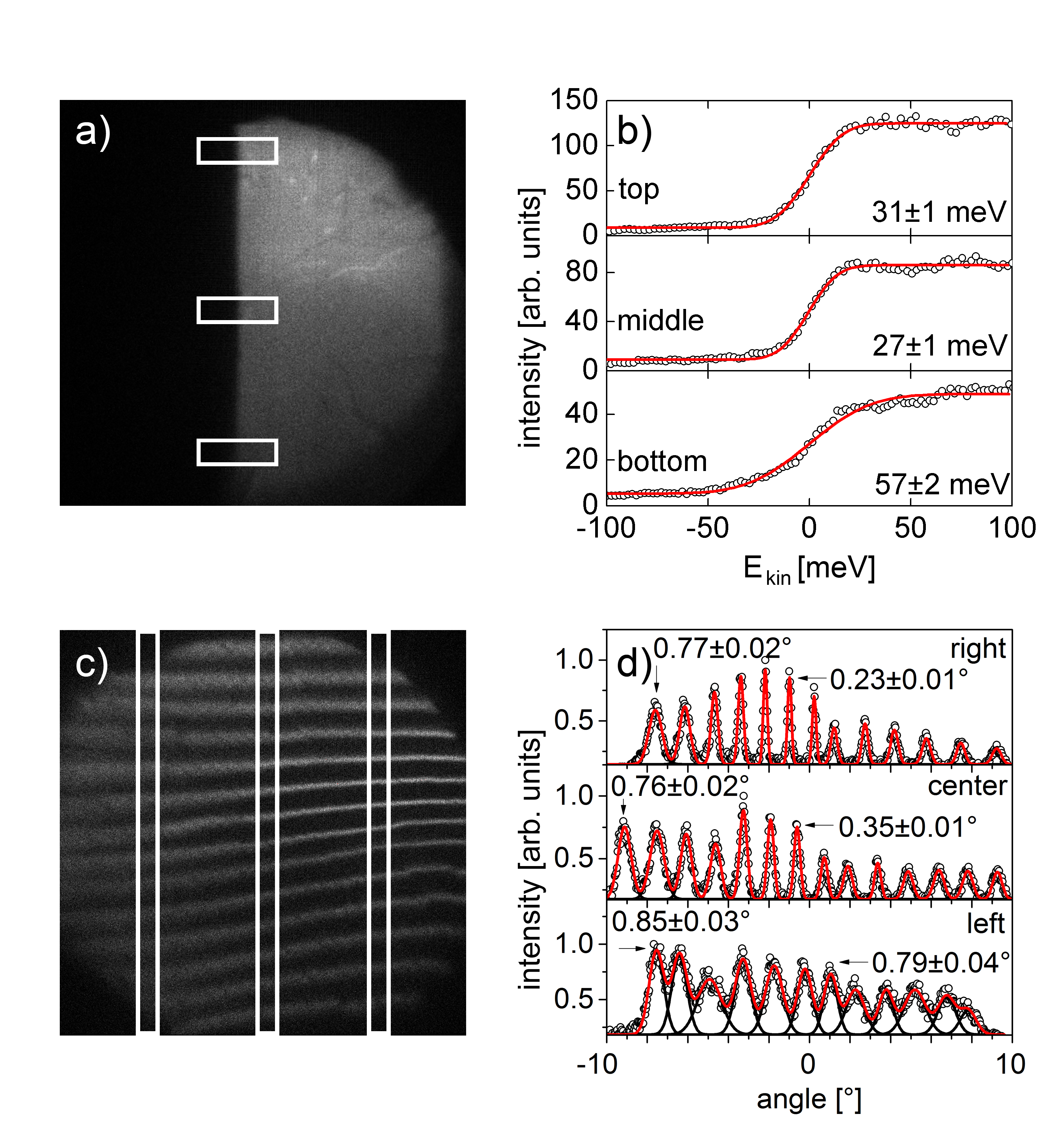}
	\caption{(a) Detector image of the low-energy cutoff of secondary eletrons produced by an electron gun. (b) Corresponding intensity profiles determined at three different emission angles as marked by the white rectangles. (c) Detector image of the angular slit array and (d) line profiles determined at three different scattering energies.}
	\label{fig:resolution}
\end{figure}

Figure \ref{fig:resolution} depicts experimental results with respect to energy and angular resolution of the imaging spin filter setup operated at a scattering energy of 10~eV. Figure \ref{fig:resolution} (a) shows an image of the low energy cutoff of the secondary electrons as a sharp boundary on the left side. The circularly shaped boundary on the right side is defined by the rim of the spin filter crystal. For this image, the electron optics have been optimized for energy resolution on the expense of angular resolution. A least squares fit (see figure \ref{fig:resolution} (b)) to the low energy cutoff using a slightly modified cumulative distribution function
\begin{equation}
I(E_{kin}) = \frac{a_0}{2} erfc \left(\frac{2 \sqrt{ln (2)}}{\Delta E_{exp}} (a_1 - E_{kin})\right) + a_2 \mathrm{,}
\end{equation}
reveals the energy resolution $\Delta E_{exp}$. Here, $erfc$ is the complementary error function, $a_0$ is the magnitude, $a_1$ defines the position of the step and $a_3$ represents a constant offset. The best value of $\Delta E_{exp} = (27 \pm 1)$~meV has been obtained in the center of the image. This value fits well to the theoretically expected value  $\Delta E_{theo} = 20$~meV for a spectrometer pass energy  of $E_{pass} = 30$~eV and slit width of 0.2~mm.\\
\\
The angular resolution has been determined for the wide angle mode using the slit array in front of the entrance lens (see figure \ref{fig:resolution} (c)). The image depicts the distortion of the electron optics and can be used for post-measurement correction. A fit to the line profiles shown in figure \ref{fig:resolution} (d) using a Gaussian function reveals a maximum angular resolution of $\Delta \alpha_{exp} = (0.23 \pm 0.01)^{\circ}$. Please note that the observed angular resolution for this measurement is limited by the spot size of 0.25~mm of the exciting electron beam on the sample. The limit of resolved data points $N$ for a circular field of 1.5~eV and $\pm10^{\circ}$ is given by
\begin{equation}
N = \frac{\pi}{4}\frac{1.5 eV}{\Delta E} \frac{20^{\circ}}{\Delta \alpha} \mathrm{.}
\end{equation}
Assuming constant minimal values $\Delta E_{exp}$ and $\Delta \alpha_{exp}$ within the detection area, $N$ equals $3.8\cdot10^3$. Considering the resolution decrease to the image boundaries, an estimated value around $N = 10^3$ is reasonable.
\subsection{Lifetime of the spin filter}
\begin{figure}
	\centering
  \includegraphics[width=0.3\textwidth]{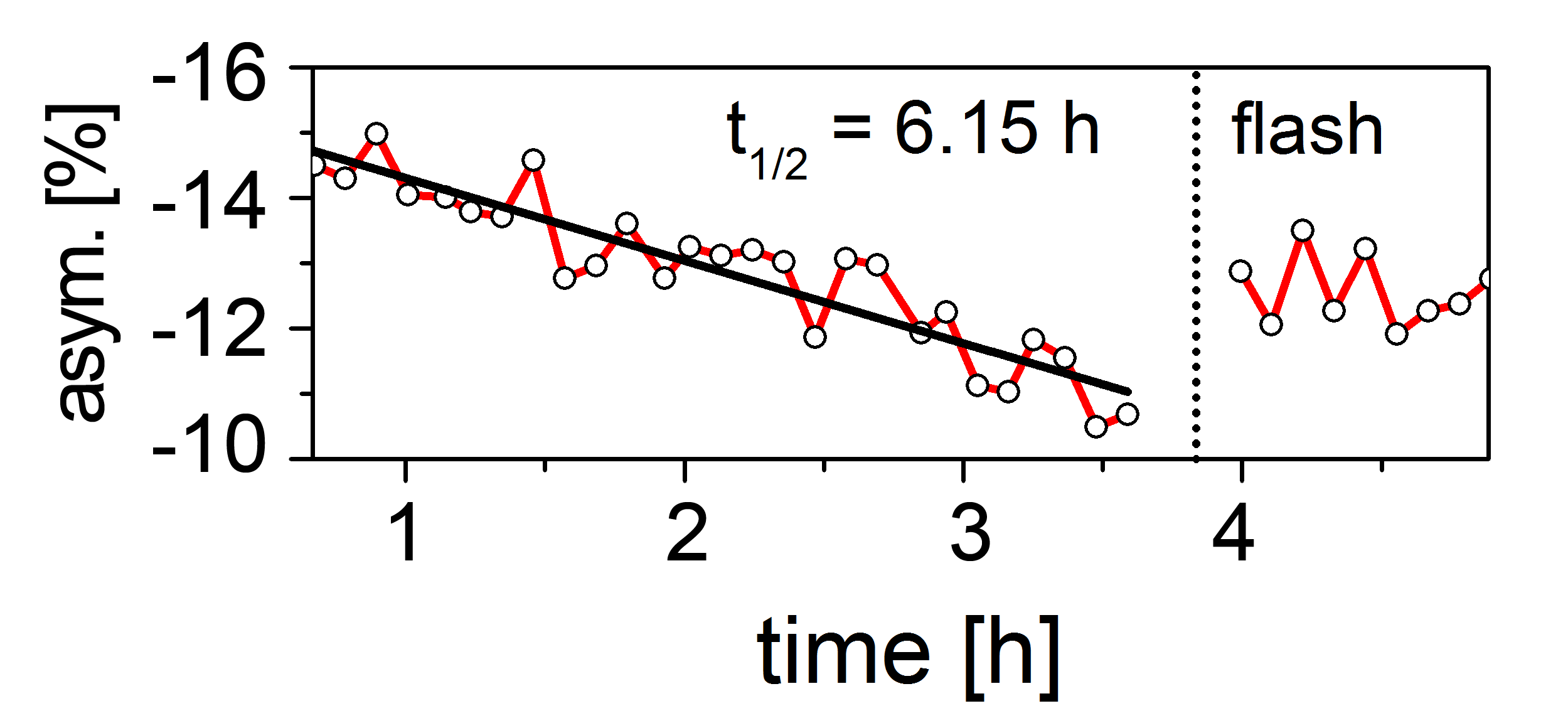}
	\caption{The lifetime of the spin filter is evaluated from a measurement series of 3 hours under constant conditions.}
	\label{fig:time}
\end{figure}

Measurements of the spin asymmetry have been performed by measuring the reflected intensity $I$ while switching the sample magnetization in opposite directions (see Refs.~\onlinecite{ELM07,KIR85}). The asymmetry is defined by equation \eqref{asymmetry} and leads to the polarization $P$ if the Sherman function $S$ is known.%\cite[p. 19]{KIR85}
\begin{equation}
A = \frac{I^{\uparrow} - I^{\downarrow}}{I^{\uparrow} + I^{\downarrow}} = S P \label{asymmetry}
\end{equation}
The multichannel efficiency is described by the figure of merit (FoM$_{2D}$) and equals $N$-times the single channel efficiency (FoM$_{single}$). It can be calculated from $S$ and the reflectivity $R$:
\begin{equation}
\mathrm{FoM_{single}} = S^2 R \mathrm{.}
\end{equation}
To analyze the asymmetry and reflectivity dependence on the spin filter temperature, a measurement series was started 7~min after the final high-temperature flash. The residual spin filter temperature of 550~K showed no significant influence on the reflected intensity and asymmetry and the values remained constant during the subsequent minutes of cooling to room temperature. Figure \ref{fig:time} shows the long-term behavior of the asymmetry at room temperature and base pressure slightly below $1\times 10^{-10}$~mbar for a period of several hours. A linear fit results in a lifetime of over 6 hours. Moreover, a high-temperature flash after 3$\frac{3}{4}$~h could recover the asymmetry from around 11\% to 13\%. A full recovery was only possible by repeating the annealing procedure and subsequently preforming a high-temperature flash.\\
\\
\subsection{Energy and Azimuthal Dependence of Ir(001)}
\begin{figure*}
	\centering
  \includegraphics[width=0.8\textwidth]{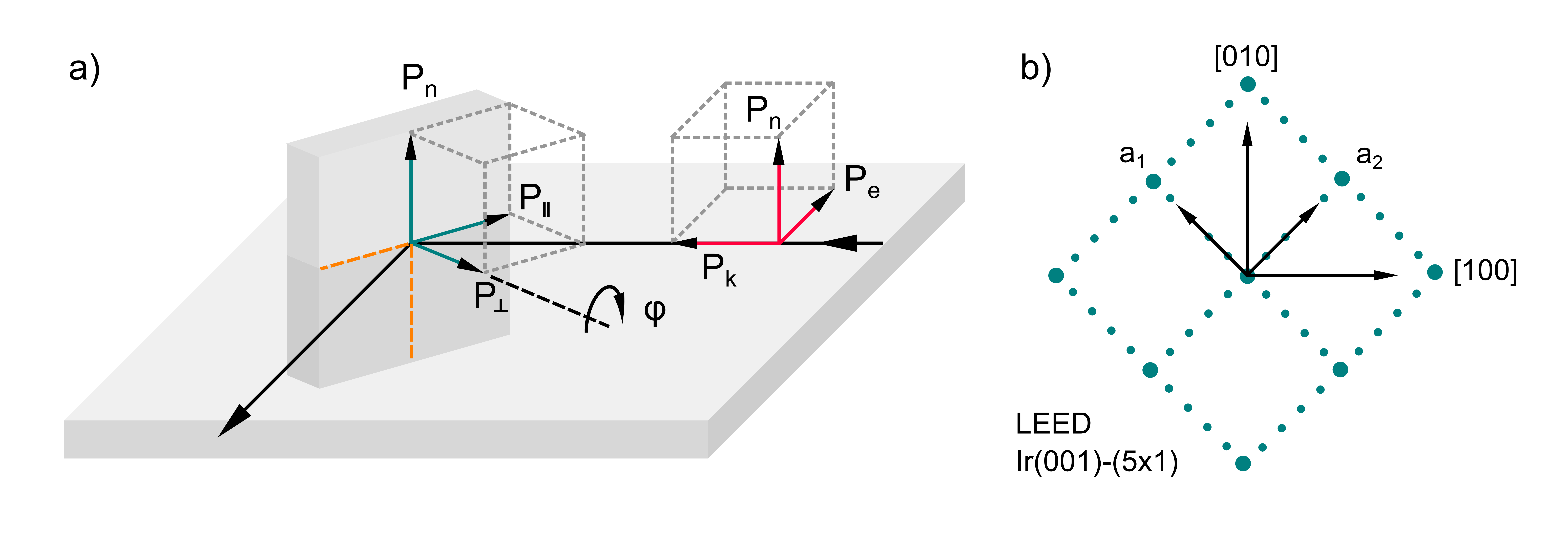}
	\caption{(a) Scattering plane of the spin filter and spin polarization component notations with respect to the incoming beam and the spin filter surface (red and green). The mirror planes of a 4-fold symmetric crystal are marked as yellow dashed lines. The spin filter can be rotated about its surface normal (angle $\varphi$). (b) Schematic LEED pattern of the Ir(001)-(5x1) surface. The used axis definition is overlayed.}
	\label{fig:geometry}
\end{figure*}

\begin{figure*}
	\centering
  \includegraphics[width=0.8\textwidth]{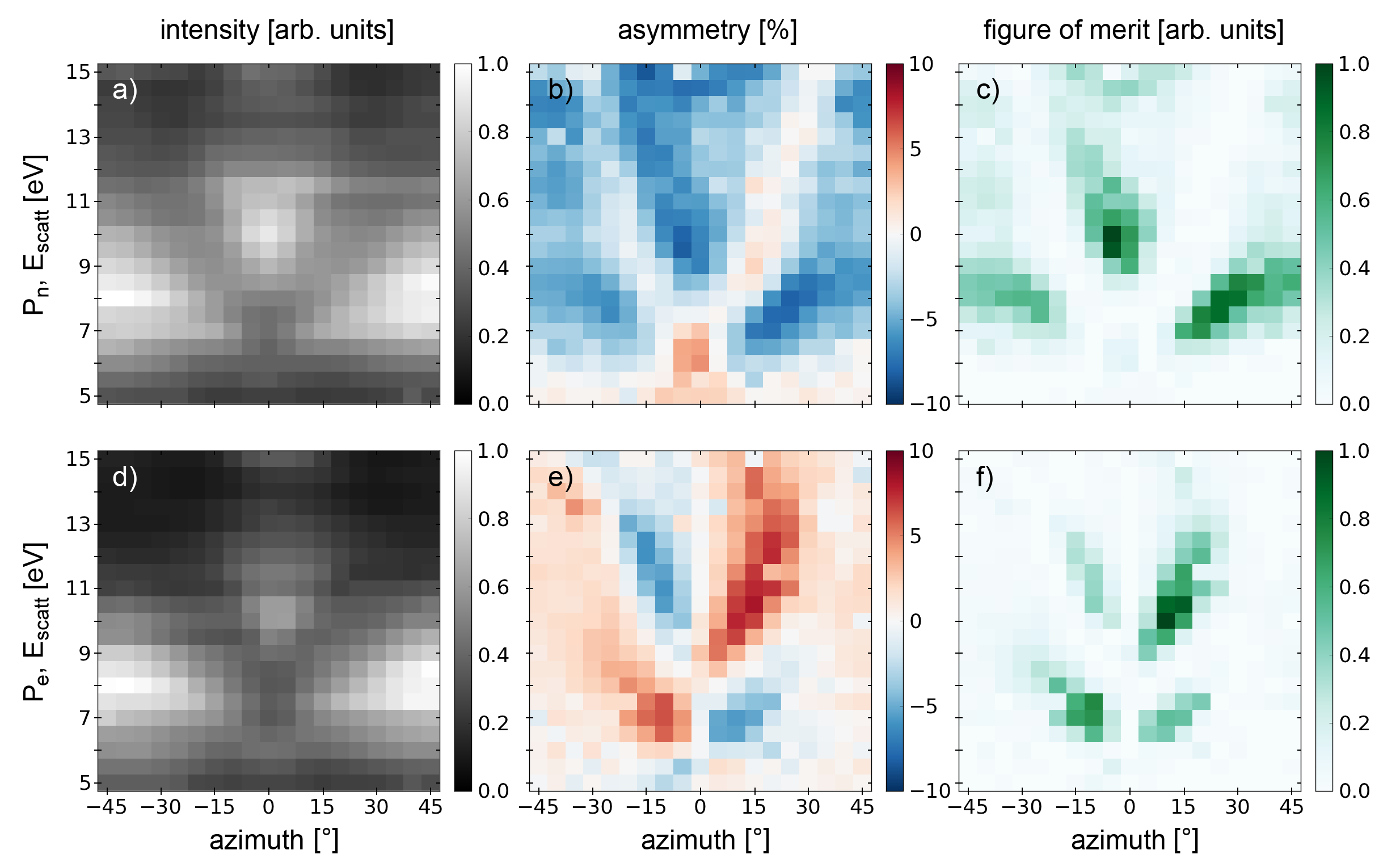}
	\caption{Intensity, asymmetry and relative FoM maps of $P_n$ (a-c) and $P_e$ (d-f) polarization components for scattering energies of 5-15~eV and a scattering angle of $45^{\circ}$ as a function of the azimuthal angle.}
	\label{fig:10ev}
\end{figure*}

\begin{figure*}
	\centering
  \includegraphics[width=0.8\textwidth]{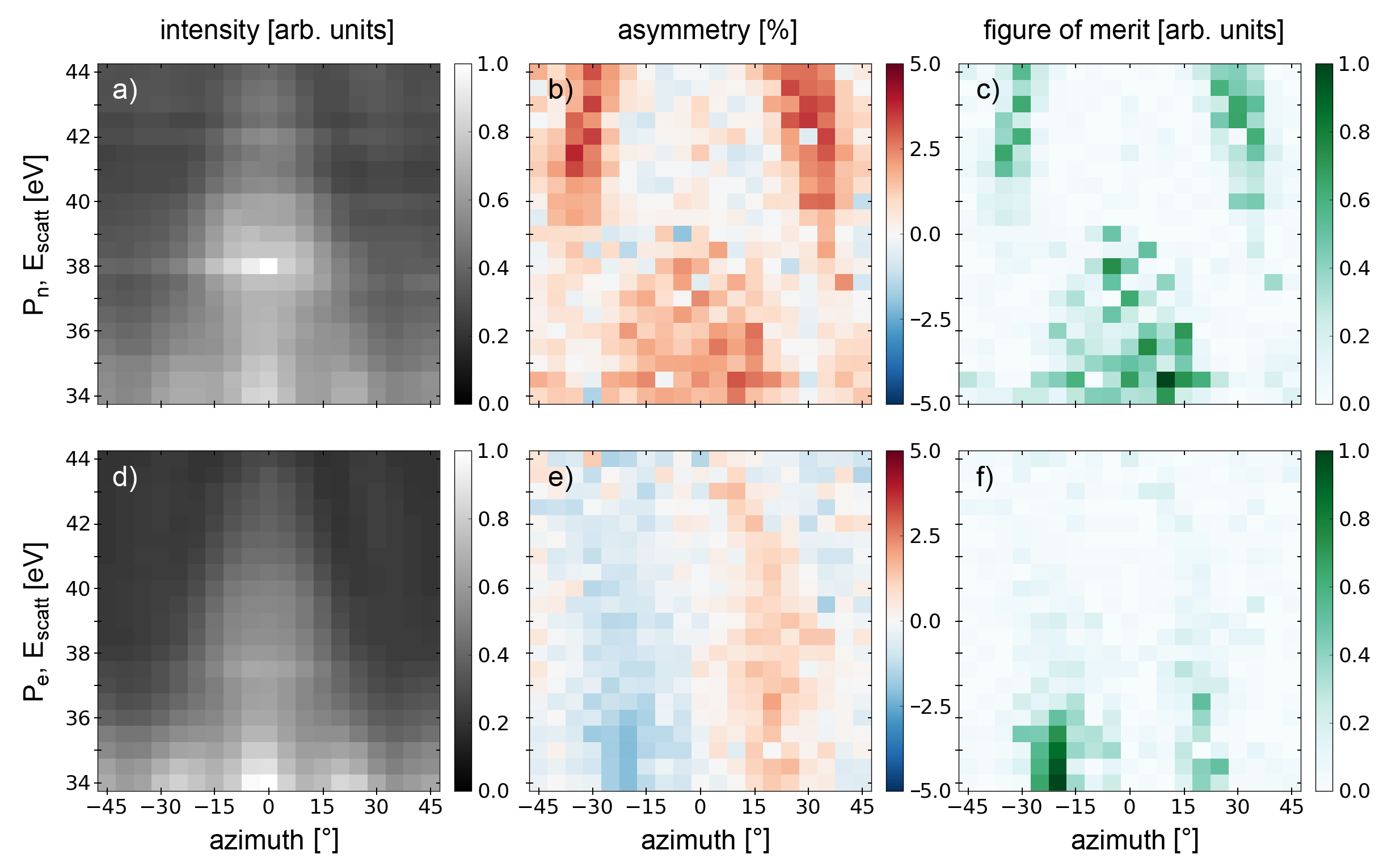}
	\caption{Intensity, asymmetry and relative FoM maps of $P_n$ (a-c) and $P_e$ (d-f) polarization components for scattering energies of 34-44~eV and a scattering angle of $45^{\circ}$ as a function of the azimuthal angle.}
	\label{fig:39ev}
\end{figure*}

A detailed characterization of the scattering energy dependence and azimuthal dependence on the intensity, asymmetry and FoM$_{single}$ has been performed. The azimuthal angle has been only varied from -45$^{\circ}$ to 45$^{\circ}$ with respect to the iridium mirror plane (100) because of the 4-fold symmetry of the Ir(001) surface. The scattering energy has been varied between 5-15~eV and 34-44~eV covering the two ranges of interesting scattering conditions identified in earlier work \cite{KUT13}. The spin polarization direction of the incoming electron beam is defined by the magnetization direction of the Fe/W(110) sample. The magnetization shows parallel or antiparallel to the easy axis given by the strong uniaxial anisotropy of the 15 monolayer thick Fe film along the $[1\overline{1}0]$ axis of the substrate crystal. The substrate crystal has been mounted in two orientations: with the ($\overline{\Gamma \mathrm{N}}$)/$[1\overline{1}0]$ or the ($\overline{\Gamma \mathrm{H}}$)/$[001]$ axis along the angular dispersive analyzer axis to obtain either a spin polarization direction within the scattering plane ($P_e$) or perpendicular to the scattering plane ($P_n$) (see figure \ref{fig:geometry}).\\
\\
Figures \ref{fig:10ev}-\ref{fig:39ev} depict maps of intensity, asymmetry and FoM for the corresponding azimuthal angle and energy ranges. Each pixel denotes one setting of the scattering conditions. A map was measured within 3-5 hours and lifetime corrections have not been applied to the raw data. The scattering energy has been derived from the kinetic energy set-point and the iridium bias voltage. Please note that the energy axis is only precise to within $\pm$ 1~eV because of the unknown work function difference of spectrometer and spin filter crystal and because of the penetration of electrical fields through the exit slit. The values have been determined from a 300 pixel wide circle, corresponding to an energy and angle interval of 0.75~eV and 12.5$^{\circ}$ (angular dispersive direction) in the center of the detector image. The level of details in the measured maps are limited by the angular spread of the incoming trajectories. Intensity and FoM$_{single}$ are given as relative values. The recording time was 5~s (34-44~eV working point) or 10~s (5-15~eV working point) per magnetization polarity. A single magnetization reversal procedure was sufficient since the values were stable. Iridium bias voltages from -5~V to 5~V in steps of 0.5~V and azimuthal angles between -45$^{\circ}$ and +45$^{\circ}$ in steps of 5$^{\circ}$ have been varied.\\
Figure \ref{fig:10ev} (a-c) depicts the results for the polarization component $P_n$, i.e. a polarization direction normal to the scattering plane, close to a scattering energy of 10~eV. An azimuthal angle of $\varphi = 0^{\circ}$ of the spin filter crystal corresponds to the $[100]$ direction lying within the scattering plane. Thus, the scattering plane is parallel to a mirror plane of the crystal. The asymmetry for this scattering geometry (figure \ref{fig:10ev} (b)) can be compared to previously published experimental and theoretical results \cite{KUT13} taken in the same geometry. The measured asymmetry shows a maximum negative value at $\varphi = 0^{\circ}$ and $E_{scatt} = 10$~eV. A polarity change occurs at $E_{scatt} = 7.5$~eV. Both observations are in fair agreement with Ref.~\onlinecite{KUT13}. The intensity (figure \ref{fig:10ev} (a)) shows a maximum value near $E_{scatt} = 10$~eV, too, leading to a significant extremal FoM for these parameters (figure \ref{fig:10ev} (c)). The intermediate intensity minimum near 10~eV as observed previously both in experiment and theory \cite{KUT13} does not appear. A possible explanation is the inherent averaging of azimuthal and polar scattering angles caused by the converging electron trajectories at the spin filter crystal. According to electron optical simulations and the known angular distribution at the exit plane the averaging angular range of azimuthal and polar scattering angles is $\Delta \varphi = 1.8^{\circ}$ and $\Delta \theta = 1.8^{\circ}$ ($E_{pass} = 30$~eV, $E_{kin} = 10$~eV, $d$ = 1~mm). The angular spread increases for an increasing distance to the optical axis and for larger entrance slits.\\
Symmetry considerations predict an even behavior $A(P_n,\varphi) = A(P_n,-\varphi)$\cite{BAU80}. This condition agrees with our experimental results within error limits. We attribute the remaining asymmetries to the macroscopic mosaic spread of the spin filter crystal and residual magnetic stray fields. Besides the maximum asymmetry at $\varphi = 0^{\circ}$ we observe a negative asymmetry extremum at $\varphi = \pm 25^{\circ}$ an $E_{scatt} = 7.5$~eV with decreasing value for decreasing scattering energy. The maximum asymmetry is again related to a maximum in intensity leading to sharp peaks in the FoM map. The regions around 7.5~eV and $\pm 25^{\circ}$ have comparable efficiency and dimensions as the scattering condition at $\varphi = 0^{\circ}$ and $E_{scatt} = 10$~eV.\\
Figure \ref{fig:10ev} (d-f) shows the corresponding results for $P_e$. In this case the mirror symmetry of the scattering experiment for $\varphi = 0^{\circ}$ and $\varphi = 45^{\circ}$ causes a vanishing asymmetry. For polycrystalline scattering targets, e.g. in a Mott detector, this condition is fulfilled independent on $\phi$ and $P_e$ cannot be measured. In contrast, if the scattering plane does not coincide with a crystal mirror plane the spin polarization component $P_e$ can lead to a finite asymmetry. In this case one expects an odd behavior of the asymmetry $A(P_e,\varphi) = -A(P_e,-\varphi)$. This is indeed confirmed by the experimental results. Two pairs of regions with large positive and negative asymmetry and FoM occur at scattering energies of  7~eV and 11~eV at  $\varphi = \pm 10^{\circ}$. The absolute maxima of the asymmetry and FoM are comparable to the maxima observed for $P_n$.\\
\\
The presence of positive and negative asymmetries for the same scattering energy represents a practical advantage for spin detection as the electron optical setting can be kept constant thus avoiding variations of the image magnification. It is also advantageous that the optimum scattering conditions for $P_e$ fall into regions of vanishing asymmetry for $P_n$ and vice versa. This allows for an independent detection of $P_n$ and $P_e$.\\
Figure \ref{fig:39ev} depicts the experimental results for the case of larger scattering energy near 39~eV. The reflected intensity shows maximum values around $0^{\circ}$ and 34~eV. The intensity monotonously decreases with increasing scattering energy.  For 34~eV a periodic intensity variation in dependence on $\varphi$ is visible.\\
For the spin polarization component $P_n$, the asymmetry shows the expected even behavior as in the case of lower scattering energy. Maximum positive asymmetry occurs for $\varphi = 0^{\circ}$, $E_{scatt} = 34$~eV and for $\varphi = \pm 30^{\circ}$, $E_{scatt} = 42$~eV. The absolute asymmetry value is a factor of two smaller than near $E_{scatt} = 10$~eV. The asymmetry maximum at $\varphi = 0^{\circ}$ coincides with a large reflected intensity and leads to a large FoM in the range of 34-39~eV. This larger energy range corresponds to the broad maximum observed in Ref.~\onlinecite{KUT13}, however, it appears at slightly lower scattering energy and thus corresponds better to the theoretical prediction (see below). The additional two parameter regions with maximum FoM are located around 42~eV and $\pm 30^{\circ}$.\\
In the case of $P_e$, the asymmetry shows the odd behavior $A(P_e,\varphi) = -A(P_e,-\varphi)$ for small values of $\varphi$ as in the case of lower scattering energy. Regions of maximum asymmetry and FoM occur at $\pm 20^{\circ}$ and $E_{scatt} = 34$ $eV$. This range of scattering energy can thus also be used for independent determination of $P_n$ and $P_e$. However, the lower absolute asymmetry values and lower reflected intensities necessitate longer acquisition times. On the other hand, electron optical distortions are less pronounced for larger scattering energies and the usable range of scattering energies is larger.\\
\\
\subsection{Spin-Polarization Manipulation with Longitudinal Magnetic Fields}
In contrast to electrical fields, magnetic fields can change the spin polarization direction. A longitudinal magnetic field $B$ parallel to the beam path causes a rotation of the transversal (perpendicular to the beam path) component of the spin polarization direction. The rotation angle is given by equation \eqref{field}. The angle increases with the time the electron is exposed to the magnetic field:
\begin{equation}
\vec{P} = \left(
\begin{matrix}
P_n cos \alpha - P_e sin \alpha\\
P_e sin \alpha + P_n cos \alpha\\
P_k
\end{matrix}
\right)
\mathrm{with~} \alpha = \frac{e}{m_e v} \int{B_z dz} \mathrm{.} \label{field}
\end{equation}
In our setup the image is already rotated by 10$^{\circ}$  without electrical current as judged from the orientation of the slit array. We attribute this rotation to a remanent magnetic field generated by the plate valve. This remanent field might also explain the remaining asymmetry observed in figure \ref{fig:10ev} (c,f). The image quality slightly decreases with larger external fields but all slits remain clearly visible in the range of applied fields.\\
\begin{figure*}
	\centering
  \includegraphics[width=0.8\textwidth]{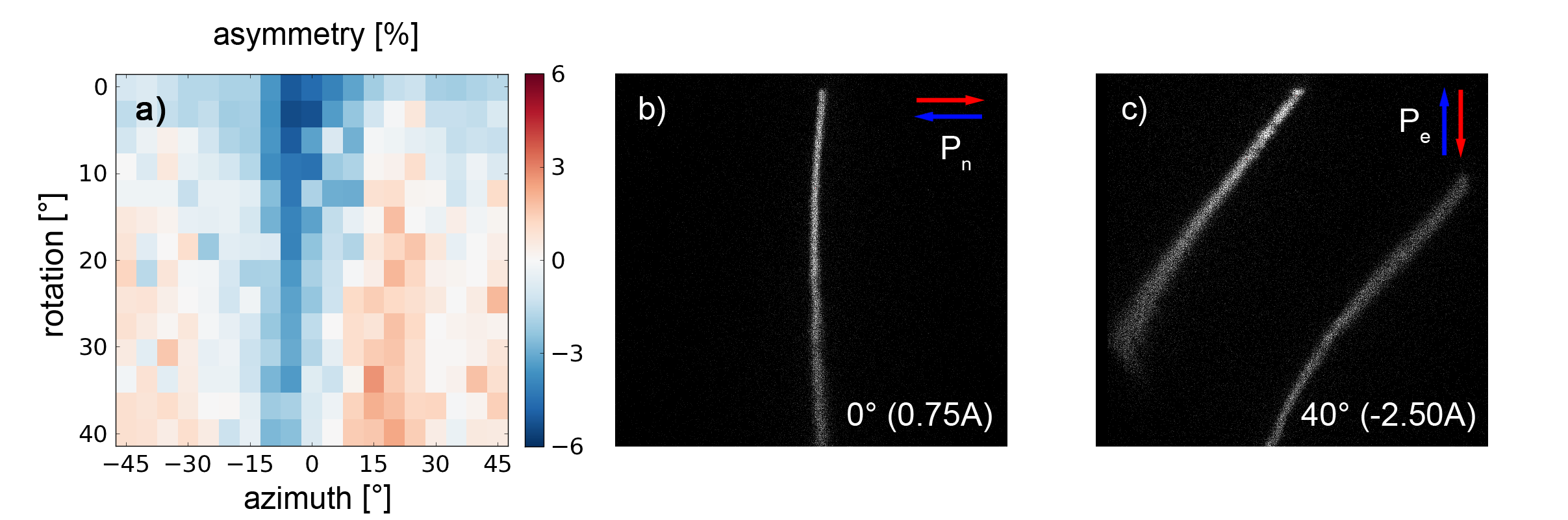}
	\caption{(a) Measured asymmetry at $E_{scatt}$ = 10~eV and azimuthal angles between $\pm$45$^{\circ}$ for different spin polarization rotation angles. (b,c) The image is rotated from 0 to 40$^{\circ}$ (spin from 0 to 80$^{\circ}$) so that the former $P_n$ component transforms into a $P_e$ component.}
	\label{fig:coil}
\end{figure*}\\
For the results shown in figure \ref{fig:coil} the scattering energy has been fixed to 10~eV while the coil current as well as $\varphi$ varies. The measured asymmetry for spin polarization $P_n$ reveals a smooth transition from a symmetric (0.75~A) to an antisymmetric (-2.5~A) behavior. The asymmetry behavior becomes symmetric at 0.75~A (0.03~mT) indicating the compensation of the longitudinal stray field component of 0.03~mT.\\
By a current of -2.5~A (0.14~mT) the asymmetry shows the odd behavior expected for polarization component $P_e$. The transversal spin polarization has rotated by 90 degrees. This field also causes a rotation of the image indicated by the direction of the slit array. The rotation roughly corresponds to an angle of $\Phi$ = $45^{\circ}$, according to equation \eqref{imagerot} as reported in Ref.~\onlinecite{EGE05}\\
\begin{equation}
\Phi = \frac{e}{2 m_e v} \int{B_z dz} = \frac{\alpha}{2} \mathrm{.} \label{imagerot}
\end{equation} 
\subsection{Theoretical Maps} 
The theoretical results support the experimental findings, i.e. a symmetric pattern for the $P_n$ component and an asymmetric pattern for the $P_e$ component with respect to $\varphi=0^{\circ}$. The origin of this particular nature of the scattering pattern can be illustrated via the component of the electron polarization perpendicular to the scattering plane. For every azimuthal angle $\varphi$ one has to calculate two scattered intensities with opposite polarization direction. This has to be done for $P_e$ and $P_n$. In the following we will address the mirror plane of the surface to an azimuthal angle of $\varphi=0^{\circ}$. Lets consider an in-plane spin polarization $P_e$. The component of $P_e$ projected on the direction perpendicular to the crystal mirror plane will be antiparallel for $\varphi<0^{\circ}$ compared to $\varphi>0^{\circ}$. This gives an inverse scattering pattern with respect to the mirror plane of the surface. Conversely, for $P_n$ the scattering will be symmetric for azimuthal angles $\varphi>0^{\circ}$ or $\varphi<0^{\circ}$. Deviations originate from the 5x1 superstructure which could not be considered in theory where we applied a (1x1) reconstructed surface. The impact of more complicated overlayers will be investigated in forthcoming work.\\
\begin{figure*}
	\centering
  \includegraphics[width=1\textwidth]{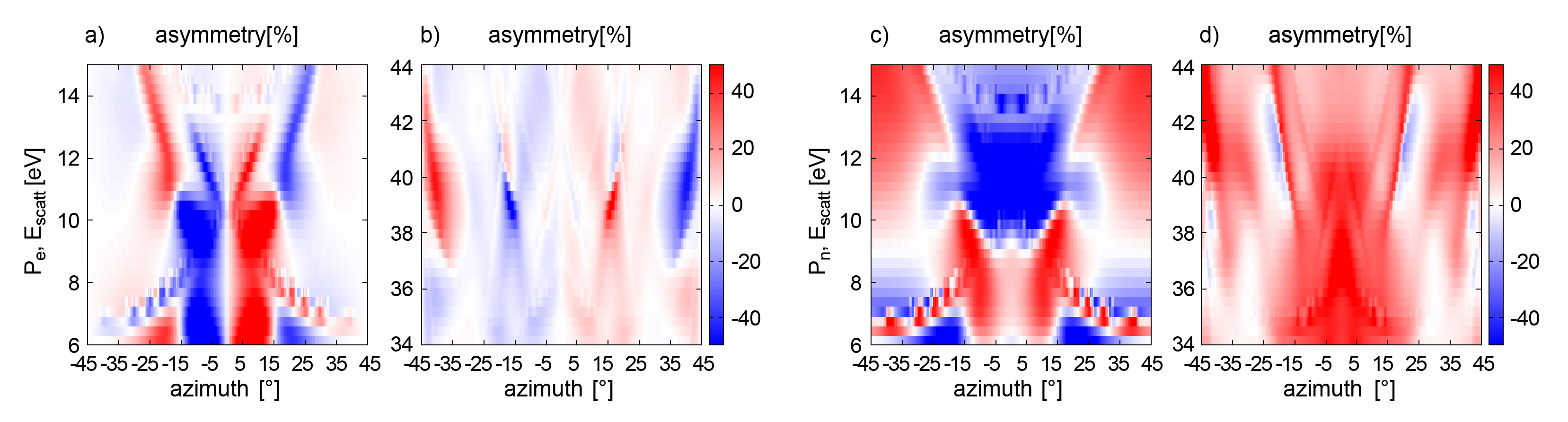}
	\caption{Theoretically determined spin-orbit induced asymmetry for the $P_e$ (a,b) and $P_n$ (c,d) component of the electron polarization in the range of 6-15~eV (a,c) and 34-44~eV (b,d).}
	\label{fig:theory}
\end{figure*}
The results of our calculations are presented in Fig. \ref{fig:theory}. For an in-plane configuration ($P_e$) our model structure nicely reproduces the scattering asymmetries for kinetic energies larger than 10~eV. For energies in the range of 6-10~eV the discrepancies between theory and experiment are larger. This can be attributed to the higher sensitivity of the electron for lower kinetic energies concerning the electronic structure at the surface, i.e. effects coming from the (5x1) superstructure. For an out-of-plane oriented polarization ($P_n$) our model reproduces the scattering behaviour for the whole energy range 6-14 eV applying azimuthal angles $\pm 20^{\circ}$. For a larger azimuthal range differences are visible which might also be related to the reconstruction. In contrast for higher kinetic energies (34-44 eV) the differences between theory and experiment are smaller and the details of the surface electronic structure become less important, i.e. deeper atomic layers contribute to the spin-orbit asymmetry.
\section{Conclusion}
Using the well-defined transversal spin polarization component of the incident electron beam, we have identified scattering parameters for the exclusive detection of the transversal spin polarization component $P_e$ parallel and $P_n$ perpendicular to the scattering plane. Assuming a spin polarization of 45\% of the incident electron beam the measured asymmetry values translate to the corresponding Sherman functions $S_n$ and $S_e$ for the normal and parallel component of the spin polarization vector. The same scattering parameters as for $P_e$ lead to a measureable asymmetry caused by the longitudinal component $P_k$. The observable asymmetries $A_1$ and $A_2$ are given by the the two disjunct scattering conditions (e.g. $E_{scatt}$ = 10~eV, $\varphi$ = 0$^{\circ}$, $\alpha$ = 0$^{\circ}$ and $E_{scatt}$ = 11~eV, $\varphi$ = $\pm 15^{\circ}$, $\alpha$ = 0$^{\circ}$). A third observable $A_3$ follows from the application of a longitudinal field rotating the transversal component by 90$^{\circ}$ (e.g. $E_{scatt}$ = 11~eV, $\varphi$ = $\pm15^{\circ}$, $\alpha$ = 90$^{\circ}$). An arbitrary spin polarization vector ($P_n$,$P_e$,$P_k$) leads to the measurable asymmetries:
\begin{equation}
\left(
\begin{matrix}
A_1\\
A_2\\
A_3
\end{matrix}
\right)
= 
\left(
\begin{matrix}
S_n & 0 & 0\\
0 & \frac{1}{\sqrt{2}}S_e & \frac{1}{\sqrt{2}}S_k\\
\frac{1}{\sqrt{2}}S_e & 0 & \frac{1}{\sqrt{2}}S_k
\end{matrix}
\right)
\left(
\begin{matrix}
P_n\\
P_e\\
P_k
\end{matrix}
\right) \mathrm{.}
\end{equation}
Inverting this equation determines the three components of the spin polarization vector from three observables. The three scattering conditions require a magnetization reversal of magnetic samples in order to generate the asymmetry via the initial beam polarization inversion. However, it is also possible to apply the vectorial spin analysis concept to non-magnetic samples. Here, six scattering conditions are needed to measure three asymmetries via the variation of scattering energies and azimuthal angles. Exemplary scattering conditions are ($E_{scatt}$ = 10~eV \& 6.5~eV, $\varphi$ = 0$^{\circ}$, $\alpha$ = 0$^{\circ}$), ($E_{scatt}$ = 11~eV, $\varphi$ = +15$^{\circ}$ \& -15$^{\circ}$, $\alpha$ = 0$^{\circ}$) and ($E_{scatt}$ = 11~eV, $\varphi$ = +15$^{\circ}$ \& -15$^{\circ}$, $\alpha$ = 90$^{\circ}$). The corresponding system of equations remains unchanged and the matrix inversion determines again the three components of the spin polarization vector.\\
In summary, we have demonstrated a concept of measuring three components of the spin polarization vector without changing the electron optical beam path. The concept is compatible with multichannel spin detection and thus allows for a significant efficiency increase of spin detection compared to classic single channel approaches.  The spin polarization component perpendicular to the scattering plane is measured if the scattering plane coincides with a crystal mirror plane. A transversal component of the spin polarization vector parallel to the scattering plane leads to an asymmetry if the scattering plane does not coincide with the crystal mirror plane. Asymmetry maps revealed scattering conditions ($E_{scatt}$,$\varphi$) for both components. The maximum observed asymmetries amount to 17\% at (10~eV, $\varphi$ = 0$^{\circ}$) for the perpendicular component and (8~eV, $\varphi$ = 15$^{\circ}$) for the out-of-plane component corresponding to a maximal Sherman function of 0.38. A longitudinal component will also lead to an asymmetry for the latter scattering condition. Switching on a longitudinal magnetic field prior to scattering distinguishes the longitudinal from the transversal component as the field rotates the transversal component but not the longitudinal component.

\begin{acknowledgments}
Funded by Stiftung Rheinland-Pfalz f{\"u}r Innovation (project 1038) and DFG (SCHO 341/9-1, BMBF (05K16UMB) and SFB/TRR 173). We would also like to acknowledge the support from the MAINZ Graduate School of Excellence (Excellence Initiative DFG/GSC 266). J{\'a}n Min{\'a}r acknowledges project CENTEM PLUS (LO1402) and VEDPMNF (CZ.02.1.01/15.003/358) of Czech ministerium MSMT.
\end{acknowledgments}

%\bibliography{bib}
%\begin{thebibliography}

%\end{thebibliography}

\end{document}